\documentclass[%
  reprint,            
  superscriptaddress, 
  amsmath,amssymb,    
  aps,                
  floatfix,           
  longbibliography    
]{revtex4-2}

\usepackage{graphicx}  
\usepackage{bm}        
\usepackage{dcolumn} 
\usepackage{subcaption}
\usepackage{graphicx}
\usepackage[percent]{overpic}
\usepackage{subcaption}
\usepackage{caption}
\usepackage[colorlinks=true,
            linkcolor=blue,
            citecolor=blue,
            urlcolor=blue]{hyperref}

\usepackage{graphicx}
\usepackage{subcaption} 
\usepackage[justification=raggedright,singlelinecheck=false]{caption}
\usepackage{microtype} 
\usepackage{comment}
\newcommand{\ve}{\varepsilon}
\newcommand{\vf}{v_\mathrm{F}}
\newcommand{\nf}{\nu_\mathrm{F}}
\newcommand{\tauee}{\tau_\mathrm{ee}}
\newcommand{\see}{\sigma_\mathrm{ee}}

\newcommand{\taui}{\tau_\mathrm{i}}

\newcommand{\bk}{\mathbf{k}}
\newcommand{\bv}{\boldsymbol{v}}
\newcommand{\bp}{\mathbf{p}}
\newcommand{\bq}{\mathbf{q}}

\newcommand{\bE}{\mathbf{E}}

\newcommand{\kf}{k_{\rm F}}
\newcommand{\ef}{\varepsilon_{\rm F}}
\newcommand{\beq}{\begin{equation}}
\newcommand{\eeq}{\end{equation}}
\newcommand{\bea}{\begin{eqnarray}}
\newcommand{\eea}{\end{eqnarray}}
\newcommand{\nn}{\nonumber}
\newcommand{\R}{\mathrm{Re}\,}
\newcommand{\I}{\mathrm{Im}\,}

\begin{document}

\title{
Conductivity of a Non-Galilean--Invariant Fermi Liquid:\\
Exact Solution of the Kinetic Equation}

\author{Tatia Kiliptari}
\email{t.kiliptari@ufl.edu}
\affiliation{Department of Physics, University of Florida, Gainesville, Florida 32611, USA}

\author{Vladimir I. Yudson}
\affiliation{Laboratory for Condensed Matter Physics, HSE University, 20 Myasnitskaya St., Moscow, 101000 Russia}

\author{Dmitrii L. Maslov}
\affiliation{Department of Physics, University of Florida, Gainesville, FL 32611, USA}

\date{\today}
\begin{abstract}

We obtain an exact expression for the conductivity 
 of a 
 non-Galilean-invariant Fermi liquid with impurities by solving  the 
kinetic equation with both screened Coulomb and $z=3$ Pomeranchuk critical interactions.
While consistent with previous asymptotic results, our solution shows that electron-electron interactions enter the conductivity solely via the quasiparticle scattering time, $\tauee$.
Accordingly, the crossovers between the collisionless and hydrodynamic regimes occur when $1/\tauee$ becomes comparable to the larger of the impurity scattering rate and the probe frequency, $\Omega$. Importantly, this crossover is governed by the quasiparticle scattering rate, rather than by a suppressed effective current-relaxation rate.
The exact solution also shows that the electron-electron correction to the \emph{dc} conductivity remains parametrically smaller than the residual conductivity throughout the Fermi-liquid regime.
For the optical conductivity, our solution also provides access to
the hydrodynamic regime, $\Omega\ll 1/\tauee$, which is inaccessible within perturbation theory. Near a $z=3$ Pomeranchuk quantum critical point, consistency between the kinetic-equation and Kubo approaches requires proper inclusion of mass renormalization within the Eliashberg approximation, which also ensures that the crossover between the collisionless and hydrodynamic regimes in the optical conductivity occurs at the Planckian scale $\Omega\sim T$.

\end{abstract}

\maketitle

\paragraph*{Introduction.} The temperature ($T$) and frequency ($\Omega$) dependencies of the electronic conductivity provide key insights into the  scattering mechanisms that  govern electron dynamics.
Traditionally, a Fermi liquid (FL) is expected to exhibit a $T^2$  \emph{dc} resistivity and Gurzhi-like optical conductivity, $\R\sigma(\Omega,T)\propto (\Omega^2+4\pi^2 T^2)/\Omega^2$ \cite{gurzhi1959mutual}, 
while the deviations are 
interpreted as signatures of non-FL physics.  
However, the $T^2$ resistivity requires either umklapp scattering \cite{landau1936collected} or compensated electron-hole 
pockets \cite{baber:1937}. 
If these conditions are not met, normal electron-electron (\emph{ee}) collisions still
affect the conductivity of a non-Galilean-invariant FL,
provided that momentum is relaxed by another mechanism, e.g., impurity scattering. 
However, the \emph{ee} contribution to the conductivity exhibits non-canonical scaling with $T$ and $\Omega$. 

In the low-temperature limit,  the dc resistivity of a 
isotropic, non-Galilean--invariant FL with impurities is given by
$\rho(T)=\rho_{\rm i}+\delta\rho(T)$,
where $\rho_{\rm i}$ is the residual resistivity and
$\delta\rho(T)\propto T^4$ (up to a logarithmic factor in two dimensions)
is independent of impurity scattering \cite{maslov:2011,pal:2012b,Kovalev:2025}. This form appears to obey
Matthiessen's rule and thus suggests a suppressed 
``current relaxation rate'' due to \emph{ee} interactions, $1/\tau_J\propto T^4\ll 1/\tauee\propto T^2$, where $\tauee$ is the quasiparticle scattering time. 
The collisionless limit of the optical conductivity
$\R\sigma(\Omega,T)\propto \max\{\Omega^4,T^4\}/\Omega^2$ \cite{rosch:2005,rosch:2006,Sharma:2021,Goyal:2023}-,
apparently reinforces this interpretation: viewed as a Drude tail,
it implies that
$1/\tau_J\propto \max\{\Omega^4,T^4\}$.
Within the same interpretation,
the low-$T$/high-$T$ crossover 
is expected to occur when
$\delta\rho(T)\sim \rho_{\rm i}$, 
implying an $\mathcal{O}(1)$ \emph{ee} contribution to the resistivity. This relatively large magnitude of $\delta\rho(T) $ would make it plausible to attribute sizable experimental variations of $\rho$, observed in a number of low-carrier-density systems \cite{sakai:2009,kamran:2015,Wang_BOS:2020,Kovalev:2025,Shilov2026},  to normal \emph{ee} collisions,  as was done, e.g., in Ref.~\cite{Kovalev:2025}. 

The motivation of this Letter is two-fold. First, we  would like to determine the actual magnitude of $\delta\rho(T)$ without relying on an interpolation between its low- and high-$T$ asymptotic forms. 
Second, we would like to explore the hydrodynamic regime, as well as the crossover between the collisionless and hydrodynamic regimes, of the optical conductivity in the absence of impurities, neither of which is accessible within perturbation theory.
To this end, we obtain an exact result for the conductivity of a non-Galilean--invariant FL by solving the kinetic equation at both zero and finite $\Omega$. This result 
shows that the low-$T$/high-$T$ crossover in the \emph{dc} resistivity and the collisionless/hydrodynamic crossover in the optical conductivity are controlled by $\tauee/\taui$ and $\Omega\tauee$, respectively, where $\taui$ is the impurity scattering time. Therefore, interpreting the previously known asymptotic results in terms of a current relaxation rate is misleading. In particular, in the \emph{dc} case we find that $\delta\rho(T)\ll \rho_{\rm i}$ throughout the temperature range of the Fermi-liquid regime.

\paragraph*{
Model.} 
As a concrete realization of a non-Galilean--invariant FL, we consider a 2D system with a linear dispersion and an energy-independent
$\taui$. While the linear dispersion is just a simple example of a non-parabolic isotropic band, the same analysis applies to a generic isotropic dispersion, with the magnitude of the effect controlled by the degree of non-parabolicity. 
Moreover, a single-valley Dirac model is relevant to a HgTe/CdTe quantum well, which is one of the systems exhibiting a sizable $\delta\rho(T)$ \cite{Kovalev:2025}.

We also employ two models of the \emph{ee} interaction.
The first model is the screened Coulomb (C) interaction, $U_\mathrm{C}(q,\omega)$, 
treated within the random-phase approximation (RPA). We assume weak coupling, which justifies the RPA treatment and, at the level of the kinetic equation, allows us to neglect Fermi-liquid renormalizations of the effective mass and the deviation of the Z-factor from unity.  
We focus on the regime of $\mathcal{E}\equiv \max\{|\Omega|,T\}\ll\Omega_{\rm p}\lesssim\ef$, where 
 $ \Omega_\mathrm{p}=\vf\kappa$ is the characteristic plasma frequency, $\vf$ is the Fermi velocity, $\kappa$ is the screening wavevector,  and $\ef$ is the Fermi energy; in this regime screening is effectively static. The extension to dynamic screening does not modify the main conclusions and will be discussed elsewhere.

The second model is a 
FL near the $z=3$ quantum critical point  (QCP), with the effective
 Hertz-Millis (HM) \emph{ee} interaction 
\cite{hertz:1976,millis:1993},
 \begin{equation}
U_{\rm HM}(q,\omega)=\frac{g}{q^2+\xi^{-2}-i\alpha\omega/q},\quad \alpha = g \nu_{\rm F}/v_{\rm F},\label{Ueff}
\end{equation}
where 
$\xi$ is the correlation length of the order parameter and $\nf$ is the density of states.  For simplicity, we assume that the instability is in the $\ell=0$ channel. The FL regime corresponds to
$\mathcal{E}
\ll \mathcal{E}_\mathrm{FL}=1/\alpha \xi^3$. We adopt the Eliashberg approximation, i.e., we assume that in the FL regime  the self-energy $\Sigma^R$ is local \cite{chubukov:2005self} and vertex corrections are small. 
In contrast to the Coulomb case, the FL renormalizations cannot be neglected close  to the QCP. The optical conductivity of the HM model has been studied by a variety of field-theoretical methods \cite{Esterlis:2021,shi2023loop,shi2024controlled,Li:2023,guo2024fluctuation,Gindikin:2024,Gindikin:2024b,
Gindikin:2025}.
Our goal is to compare these results with our result obtained within the KE approach.

\paragraph*{Main results: General.}
For the two models specified above, we solve the kinetic equation (KE)
exactly--by mapping 
it onto an inhomogeneous Legendre-type differential equation.
The resulting conductivity is given by
\begin{subequations}
\bea
\sigma(\Omega,T)&=&\sigma_\mathrm{i}(\Omega)+\see(\Omega,T),\label{scalingF}\\
\see(\Omega,T)&=& e^2\ef \taui(\Omega)
\Phi\left(\Omega,T\right) 
\mathcal{F}\!\left(\frac{\tau_{\rm ee}(\Omega,T)}{\taui(\Omega)}
\right),\label{scalingF2}
\eea
\end{subequations}
where $\taui(\Omega)=\taui/(1-i\Omega\taui)$, $\sigma_\mathrm{i}(\Omega)=e^2\nu_F v_F^2\taui(\Omega)
$ is 
the residual conductivity,
$\nu_F=\ef/2\pi v_F^2$ is the density of states per spin, 
and $\Phi(\Omega,T)=
(T^2+\Omega^2/
4\pi^2)
/\ef^2
$. (We set $k_B=1$ and $\hbar=1$, except in the plots.)

The quasiparticle scattering rate, $1/\tauee(\Omega,T)$, 
related to
the imaginary part of the on-shell 
self-energy, is given by
$\tauee^{-1}=
a_{\rm M}\mathcal{E}^2 L_{\rm M}
(\mathcal{E}),
$
where ${\rm M}={\rm C,HM}$ indicates the model and 
$ L
_{\rm M}
(\mathcal{E})\equiv \ln\left(
\Lambda_{\rm M}/\mathcal{E}\right)$. 
Whether the FL parameters
are taken as bare or renormalized also depends on the model, i.e., C  vs. HM.  
The function $\mathcal{F}(x)$ is expressed in terms of the associated Legendre functions, with the following asymptotic limits $\mathcal F(|x|\gg 1)=-4\pi/15x$ and $\mathcal F(|x|\ll 1)=-\pi/6+\pi x/8$.

\paragraph*{Main results: Coulomb interaction.}
For this model, all Fermi-liquid parameters are bare, with $a_{\rm C}\sim 1/\ef$ and $\Lambda_{\rm C}\sim \Omega_{\rm p}$. In the \emph{dc} limit ($\Omega=0$), Eqs.~\eqref{scalingF} and \eqref{scalingF2} yield the known low-$T$
behavior $\sigma_{ee}(0,T)\propto T^4\ln T$ 
\cite{pal:2012b,Sharma:2021,Kovalev:2025}
and the high-$T$ saturation of the conductivity,
up to a Sommerfeld-type correction of order $(T/\ve_{\rm F})^2$ \cite{pal:2012,Kiliptari2025}. Beyond reproducing these limits, the exact solution also establishes that the low-$T$/high-$T$ crossover occurs when $\gamma_T \sim 1/\taui$, with 

\bea
\gamma_T\equiv 1/\tauee(0,T).\label{gamma}
\eea

Moreover,  since $|\see|/\sigma_{\rm i}(0)\lesssim (T/\ef)^2\ll 1$, the \emph{ee} contribution remains small, rather than order one, in the degenerate regime. A comparison of $\see$ with quantum-interference corrections  
is discussed in the Supplemental Material \cite{SMI}, which includes Refs. \cite{altshuler:1985,Zala:2001,gornyi:2004,Paul:2005,levitov:2019,Sharma:2021,Goyal:2023,rammer:1986,mahan:book,rammer:1986,lifshitz:1980}.

The physical origin of both the $T^4\ln T$ scaling and the low-$T$/high-$T$ 
crossover can be understood by noting that a normal \emph{ee} collision relaxes the current provided that the total velocity of the two electrons changes in the collision. This change is
\bea
\Delta\bv=\bv_\bk + \bv_\bp
- \bv_{\bk-\bq} - \bv_{\bp+\bq}.\label{Deltav}
\eea
For a general isotropic, non-parabolic dispersion, 
$\Delta\bv\neq 0$. Upon projection onto the Fermi surface (FS), however, $\bv_\bk\propto \bk$ 
and $\Delta\bv=0$ by momentum conservation, as in a Galilean-invariant system.
The leading FS contribution therefore cancels, and the conductivity is controlled by states close to, but not exactly on, the FS. Thermal smearing then supplies a factor $\propto T^2$, represented by $\Phi(0,T)$ in Eq.~\eqref{scalingF2}. Combined with the quasiparticle rate $\gamma_T\propto T^2\ln T$, this yields the $T^4\ln T$ scaling. Since $\mathcal{F}$ depends only on $1/\gamma_T\taui$, the crossover occurs when $\gamma_T\taui\sim 1$, while no separate ``current relaxation rate'' appears in the result.

For the optical conductivity,
we focus on the impurity-free limit, $\taui\to\infty$, in which
the first term in Eq.~\eqref{scalingF} is singular, with a real part
$\propto\delta(\Omega)$. The regular absorption is then determined by
the second term. This term reproduces the previous collisionless result
$
\R\see(\Omega,T)
\propto \mathcal{E}^4 L_{\rm C}(\mathcal{E})/\Omega^2$
\cite{Sharma:2021,Li:2023}, 
while specifying the collisionless condition
as $\Omega\gg\gamma_T$. It also predicts a finite hydrodynamic limit,
$\R\see
\sim e^2T^2/\gamma_T\ef$,  
for $\Omega\ll\gamma_T$.
(In a gapless Dirac system, interband electron-hole processes generate an additional  
contribution to the optical conductivity in the hydrodynamic limit \cite{Mueller:2008b}, which is not accounted for in our single-band model.)

\paragraph*{Main results: Hertz-Millis interaction.}
The final result for the conductivity is still given by
Eqs.~\eqref{scalingF} and \eqref{scalingF2}, but with the bare
parameters replaced by their renormalized counterparts,
$\vf^*/\vf=\ef^*/\ef=\nf/\nf^*=Z$, where
$Z=(1-\partial_\omega\R\Sigma^R(\omega))^{-1}$ is the quasiparticle
residue
 \footnote{Within the Eliashberg approximation, the $Z$ factor is tied
to the mass renormalization, so its appearance does not violate gauge
invariance.}.
Close enough to the QCP, $Z\sim \vf/g\xi\ll 1$~\cite{Pimenov:2022}.
In addition, $1/\tauee$ is replaced by the renormalized rate
$1/\tauee^*(\mathcal{E})\sim Z|\I\Sigma^R(\mathcal{E})|$, where
$|\I\Sigma^R(\mathcal{E})|\sim
\ef g^2(\xi/\vf)^4\mathcal{E}^2 L_{\rm HM}$ and $L_{\rm HM}=\ln(\mathcal{E}_{\rm FL}/\mathcal{E})$~\cite{Pimenov:2022}.
The distinction between $|\I\Sigma^R|$ and $Z|\I\Sigma^R|$ is
consistent with the ``resilient quasiparticle''
concept~\cite{Deng:2013}. 

The effect of these renormalizations is most transparent for the
impurity-free optical conductivity, whose asymptotic limits are captured
by the interpolation formula
\bea
\R\see(\Omega,T)\sim e^2\frac{\mathcal{E}^2}{Z\ef}
\frac{\tauee^*(\mathcal{E})}
{\Omega^2\tauee^{*2}(\mathcal{E})+1}.
\label{inter}
\eea
For $\Omega\gg T$, $\tauee^*(\mathcal{E})
\approx \tauee^*(\Omega,0)\gg 1/\Omega$, as required for a FL.
Accordingly, $\R\see(\Omega,0)\sim
e^2 g^2(\xi/\vf)^4\Omega^2 L_{\rm HM}(\Omega)$, with all FL
renormalizations canceling out, in agreement with
Ref.~\cite{Gindikin:2025}. In the hydrodynamic regime,
$\Omega\ll 1/\tauee^*(0,T)\ll T$, one finds
$\R\see(\Omega\to 0,T)\sim
e^2\vf^4/Z^2g^2\ef^2\xi^4L_{\rm HM}(T)$; here the factor
$Z^{-2}$ remains. Within the FL regime, the $Z$ factor does not depend on $T$ and $\Omega$.

For $\mathcal{E}\gtrsim \mathcal{E}_{\rm FL}$, the system enters a non-FL regime, where the 
KE 
does not apply.
We can nevertheless obtain the qualitative behavior in this regime by
replacing $\xi^{-1}$ with its dynamic counterpart,
$\xi^{-1}\to \xi^{-1}(\mathcal{E})\sim
(g/\vf)(\mathcal{E}/\mathcal{E}_0)^{1/3}$, where
$\mathcal{E}_0\sim g^2/\ef$, as in
Refs.~\cite{chubukov:2017,Li:2023,Gindikin:2024}.
Accordingly, $Z\to Z(\mathcal{E})\sim
(\mathcal{E}/\mathcal{E}_0)^{1/3}$.
For the $z=3$, $d=2$ QCP, $|\I\Sigma^R(\mathcal{E})|
\propto\mathcal{E}^{2/3}$ and
$Z(\mathcal{E})\propto\mathcal{E}^{1/3}$; hence the renormalized
rate $1/\tauee^*$ becomes Planckian: $1/\tauee^*\sim\mathcal{E}$, 
while $|\I\Sigma^R(\mathcal{E})|$ itself is super-Planckian.
Consequently, we obtain
$\R\see(\Omega,0)\propto |\Omega|^{2/3}$, in agreement with
Ref.~\cite{Gindikin:2025}, and
$\R\see(\Omega\to 0^+,T)\propto T^{2/3}$, with a crossover between
the two limits at $\Omega\sim T$.

We now outline the derivation of the results summarized above, with further details provided in Appendix~\ref{app:reduction}.

\paragraph*
{Kinetic equation with Coulomb interaction.} 
The semiclassical KE,
linearized near equilibrium, reads
\begin{equation}
    -i\Omega\delta f_{\mathbf{k}}-e\mathbf{E}\cdot\bv_\bk
    n'_\bk
    = -I_{\mathrm{ee}}[f_{\mathbf{k}}]
    -\frac{\delta f_\bk
    }{\taui},
    \label{KE}
\end{equation}
where \(n_{\mathbf{k}}\equiv n(\ve_\bk)\) is the Fermi function, $n'_\bk\equiv \partial_{\ve_\bk} n_\bk$, $\boldsymbol{v}_\mathbf{k}=\partial_\bk\ve_\bk$, \(I_{\rm ee}\)
is the \emph{ee} collision integral,
and the last term describes  $ei$ scattering.
The deviation from equilibrium, $\delta f_{\mathbf{k}}$, is chosen as
$\delta f_{\mathbf{k}}
= f_{\mathbf{k}} - n_{\mathbf{k}}
\equiv-T n'_{\mathbf{k}}\, g_{\mathbf{k}}$.
The linearized \emph{ee} collision integral is  given by \cite{abrikosov:book}
\bea
\!\!\! I_\mathrm{ee}[f_{\mathbf{k}}]=
\int_{\bp,\bk'}\!\!\!\!\!
W_{\mathbf{k},\mathbf{p};
\mathbf{k'}\mathbf{p'}}
(g_{\mathbf{k}}+g_{\mathbf{p}}-g_{\mathbf{k'}}-g_{\mathbf{p'}})
n_\mathbf{k}n_\mathbf{p}
\bar n_{\bk'} \bar n_{\bp'},\label{Ieelin}
\eea
 where 
 $\int_\bk\equiv \int d^2k/(2\pi)^2$
and $\bar n_\bk\equiv 1-n_\bk$. 
For $T\ll \Omega_{\rm p}$, the scattering kernel is expressed via
the statically screened Coulomb potential as
$W_{\mathbf{k},\mathbf{p};
\mathbf{k'}\mathbf{p'}}=2\pi
\delta(\mathbf{k}+\mathbf{p}-\mathbf{k'}-\mathbf{p'})\delta(\varepsilon_\mathbf{k}+\varepsilon_\mathbf{p}-\varepsilon_\mathbf{k'}-\varepsilon_\mathbf{p'})W(q)$, where 
$W(q)=2\pi |U(q=|\bk-\bk'|,\omega=0)|^2=
(2\pi)^3e^4/(q+\kappa)^2$. Since a weak Coulomb interaction implies that $\kappa\ll \kf$, we neglect exchange processes.

 In linear response, $\delta f_\mathbf{k}$ must be proportional to $\bv_\bk\cdot\mathbf{E}$. 
Parametrizing $g_\mathbf{k} = -(e E/T)(\boldsymbol{v}_\mathbf{k}\cdot \hat{\mathbf{e}}) \big[\taui(\Omega) + F(\varepsilon_\mathbf{k})\big]$
with $\hat{\mathbf{e}}=\mathbf{E}/E$,
we rewrite Eq.~\eqref{KE}
as
\bea
\frac{F(\ve_{\bk})}{\taui(\Omega)}(\boldsymbol{v}_{\bk}\cdot\hat{\mathbf{e}})+S[F]=-I(\bk),\,\mathrm{where}\label{KE2}
\eea
\begin{widetext}
\begin{subequations}
\begin{align}
&I(\mathbf{k}) = \taui(\Omega) \int_{\mathbf{p}\mathbf{q}\omega} W(q)\frac{\bar n(\ve_\bk-\omega)}{\bar n(\ve_\bk)}
\, n(\ve_{\mathbf{p}}) 
\bar n(\ve_\bp+\omega)
\left(\Delta\bv
\cdot \hat{\mathbf{e}}\right)\,  
\delta(\ve_{\mathbf{k}-\mathbf{q}} - \ve_{\mathbf{k}} + \omega) 
\delta(\ve_{\mathbf{p}+\mathbf{q}} - \ve_{\mathbf{p}} - \omega),
\label{source} \\[1mm]
&S(\mathbf{k}) = \int_{\mathbf{p}\mathbf{q}\omega} W(q)
\frac{\bar n(\ve_\bk-\omega)}{\bar n(\ve_\bk)}
\, n(\ve_{\mathbf{p}}) 
\bar n(\ve_\bp+\omega)
\Big\{
\bv_\bk [F(\ve_{\mathbf{k}})-F(\ve_{\mathbf{k}}-\omega)] 
- (\bv_{\bk-\bq}-\bv_\bk) F(\ve_{\mathbf{k}}-\omega) \notag\\
& + \bv_\bp [F(\ve_{\mathbf{p}})-F(\ve_{\mathbf{p}}+\omega)] 
- (\bv_{\bp+\bq}-\bv_\bp) F(\ve_{\mathbf{p}}+\omega)
\Big\}\cdot\hat{\mathbf{e}} 
\;\delta(\ve_{\mathbf{k}-\mathbf{q}} - \ve_{\mathbf{k}} + \omega) 
\delta(\ve_{\mathbf{p}+\mathbf{q}} - \ve_{\mathbf{p}} - \omega),\label{Sterm}
\end{align}
\end{subequations}
\end{widetext}
where $\Delta\bv$ is defined in Eq.~\eqref{Deltav}.

The function $F$ can be decomposed into the  even and odd in energy parts: $F(\ve_\bk)=F_\mathrm{e}(\ve_\bk)+F_\mathrm{o}(\ve_\bk)$, where $F_\mathrm{e}(-\ve_\bk)=F_\mathrm{e}(\ve_\bk)$ and $F_\mathrm{o}(-\ve_\bk)=-F_\mathrm{o}(\ve_\bk)$. 
Recalling that for Dirac fermions $\nu(\ve_\bk)=\nu_F\left(1+\ve_\bk/\ef\right)$ and $v_F=\mathrm{const}$,  
the \emph{ee} part of the conductivity becomes
\bea
\see(\Omega,T)=
e^2\nu_F v_F^2
\!\int 
d\ve_\bk (-n'(\ve_\bk))\!\!\left[F_\mathrm{e}(\ve_\bk)+\frac{\ve_\bk}{\ef}F_\mathrm{o}(\ve_\bk)\right].\label{cond}
\eea
The relation between $F_\mathrm{e}$ and $F_\mathrm{o}$ follows from momentum conservation, 
$\int_\mathbf{k} I_{ee}[f_\bk] \, \mathbf{k} = 0$, 
which   implies that 
\begin{align}
\int d\varepsilon
\, F_\mathrm{e}(\ve)
\, n'(\ve)
= -2 \int d\varepsilon
\, \frac{\varepsilon
}{\varepsilon_{\rm F}} \, F_\mathrm{o}(\ve
)\, n'(\ve)
. 
\label{constraint}
\end{align}
Therefore, the conductivity can be expressed via either $F_\mathrm{e}$ or $F_\mathrm{o}$, and we will focus on the latter. 
Equation~\eqref{constraint}
also implies that
$
\left\vert F_\mathrm{o}(\varepsilon
)\right\vert\gg \left\vert F_{e}(\varepsilon)\right\vert$,
which allows one to simplify the KE.

Next, we expand $\Delta\bv$ in Eq.~\eqref{source} near the FS; in Eq.~\eqref{Sterm}, 
only the $\bv_{\bk}[F(\ve_k)-F(\ve_k-\omega)]$ term contributes to the odd sector to logarithmic accuracy.
The angular integrations in Eqs.~\eqref{source} and \eqref{Sterm} are carried out 
with the help of the energy-conserving $\delta$-functions. The integral over $q$ is evaluated as 
$\int_{T/\vf}^{\infty} (q \, dq / 2\pi) \, W(q,0)/q^2 = L_{\rm C}(T)/4\nu_{\rm F}^2$.  
Finally, integrating over the energy $\ve_\bp$ and singling out the odd in energy part,
we arrive at the integral equation for  
$\mathcal{H}_\mathrm{o}(\xi)\equiv T F_\mathrm{o}(\xi)/\cosh(\xi/2)$:
\begin{equation}
2\!\! \int d\zeta \, \mathcal{M}_\mathrm{e}(\xi-\zeta) \mathcal{H}_\mathrm{o}(\zeta)
- \left[ \pi^2\lambda + \mathcal{N}_\mathrm{e}(\xi) \right] \mathcal{H}_\mathrm{o}(\xi)
= 
\mathcal{Y}(\xi,\lambda)
,\label{KEmain}
\end{equation}
where $\xi=\ve_\bk/T$,
$\lambda=1/\gamma_T\taui(\Omega)$,
and $\gamma_T\equiv \pi T^2 L_{\rm C}(T)/4\ef$.
Furthermore, $\mathcal{M}_\mathrm{e}(\xi)=\xi/\left(2\sinh(\xi/2)\right)$, $\mathcal{N}_\mathrm{e}(\xi)=\xi^2+\pi^2$, and  
$\mathcal{Y}(\xi,\lambda)=-(8/3\pi\lambda L_{\rm C}) \xi(\xi^2+\pi^2)/\cosh(\xi/2)$. 
 
As in the previous work \cite{abrikosov:1959,Sykes, *Sykes_2,Wilkins,lyakhov:2003,li:2018},
we Fourier-transform the integral equation \eqref{KEmain} into a differential one.  Taking into account that $\xi^2 \to -d^2/dt^2$ and introducing  $s = \tanh(\pi t)$, we obtain
\begin{align}
\left\{ \frac{d}{ds} \left( (1-s^2) \frac{d}{ds} \right) + \left[ 2 - \frac{\beta^2}{1-s^2} \right] \right\} \mathcal{H}_\mathrm{o}(s) 
= -i V_\mathrm{o}(s),
\label{diffeq2}
\end{align}
where $V_\mathrm{o}(s)=(32 \pi / \lambda L_{\rm C}) \, s \sqrt{1-s^2}$ and  $\beta = \sqrt{1 + \lambda
}$.
The fundamental solutions of Eq.~\eqref{diffeq2} can be chosen as the associated Legendre functions, 
\begin{align}
&P^{\pm\beta}_1(s)= (s\mp\beta) \left( \frac{1+s}{1-s} \right)^{\pm\beta/2}. 
\label{psib}
\end{align}
Constructing the Green's function of Eq.~\eqref{diffeq2} 
as
\begin{align}
\mathcal{R}_\beta(s,s') = \mathcal{C_\beta} \,
\begin{cases}
P^{-\beta}_1(s) P^{\beta}_1(s'), & s > s',\\
P^{-\beta}_1(s') P^{\beta}_1(s), & s < s',
\end{cases}
\label{Gf2}
\end{align}
where $\mathcal{C_\beta}=1/2\beta(\beta^2-1)$,
we obtain its solution
\begin{align}
\mathcal{H}_\mathrm{o}(s) &= -i \int_{-1}^1 ds' \, \mathcal{R}(s,s') V_\mathrm{o}(s').
\label{GRV}
\end{align}
Using the constraint (\ref{constraint}) and substituting Eq.~\eqref{GRV} into Eq.~\eqref{cond}, we obtain the scaling form of $\see(\Omega,T)$, announced in Eq.~\eqref{scalingF2}, with $\Phi(0,T)=T^2/\ve_\mathrm{F}^2$, $\tauee^{-1}(\Omega,T)=\gamma_T $
and 
\bea
\mathcal{F}(\lambda)\equiv
2\pi\int_0^1 ds s\int_{-1}^1 ds' s'\,\frac{
\sqrt{1-s'^2}
}{\sqrt{1 - s^2}}\mathcal{R}_\beta(s,s'), \label{functionF}
\eea
where 
$\beta$ is 
related to $\lambda$ as specified after Eq.~\eqref{diffeq2}.
The double integral above defines the exact solution and
can be evaluated to any desired numerical accuracy. The asymptotic limits of $\mathcal F$ have been specified after Eq.~\eqref{scalingF2} and are derived in Appendix~\ref{app:reduction}.
We now discuss the various regimes, starting with the \emph{dc} limit.
 At low temperatures 
 ($1/\gamma_T\gg\taui$),
 we find $\sigma(0,T)-\sigma_i(0)=-4\pi e^2T^2\taui^2\gamma_T/15\ve_F\propto T^4|\ln T|$,  in agreement with 
Refs.~\cite{Sharma:2021,Kovalev:2025,Kiliptari2025}. At high temperatures 
($1/\gamma_T\ll\taui$ 
but still $T\ll \Omega_{\rm p}$), $\gamma_T$ drops out of the result, yet the conductivity continues to decrease: $\sigma(0,T)=\sigma_i(0)-\pi e^2 T^2\taui/6\ve_F$. The $O(T^2)$ term is the Sommerfeld-like correction to the Drude conductivity, consistent with the spectral analysis of the collision operator \cite{pal:2012b,Kiliptari2025}.
 
\begin{figure}[h!]
    \centering
    \begin{subfigure}{\linewidth}
        \centering
        \includegraphics[width=\linewidth]{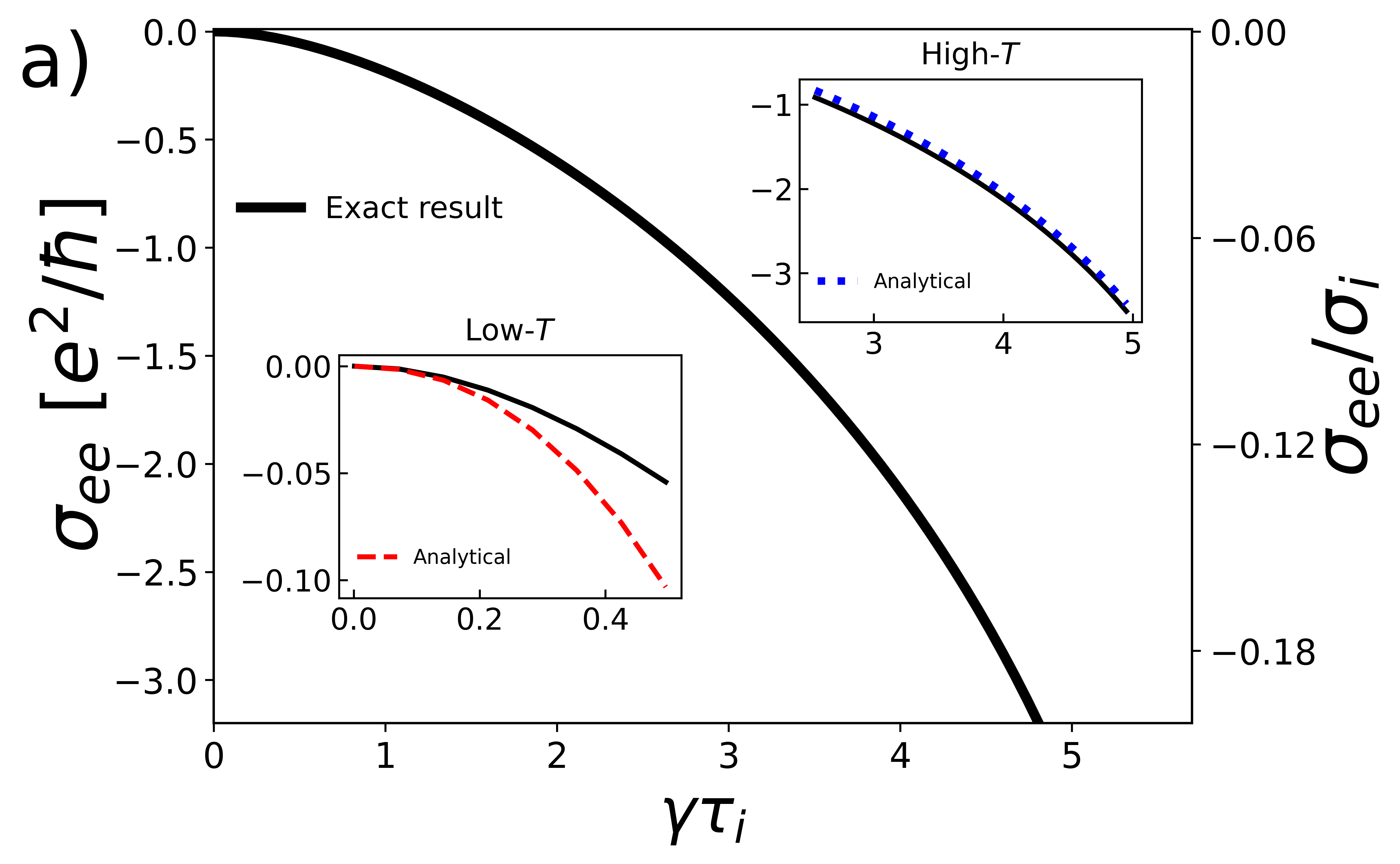}
        \label{dc}
    \end{subfigure}
    \begin{subfigure}{\linewidth}
        \centering
        \hspace*{-1.3cm}
\includegraphics[width=0.86\linewidth]{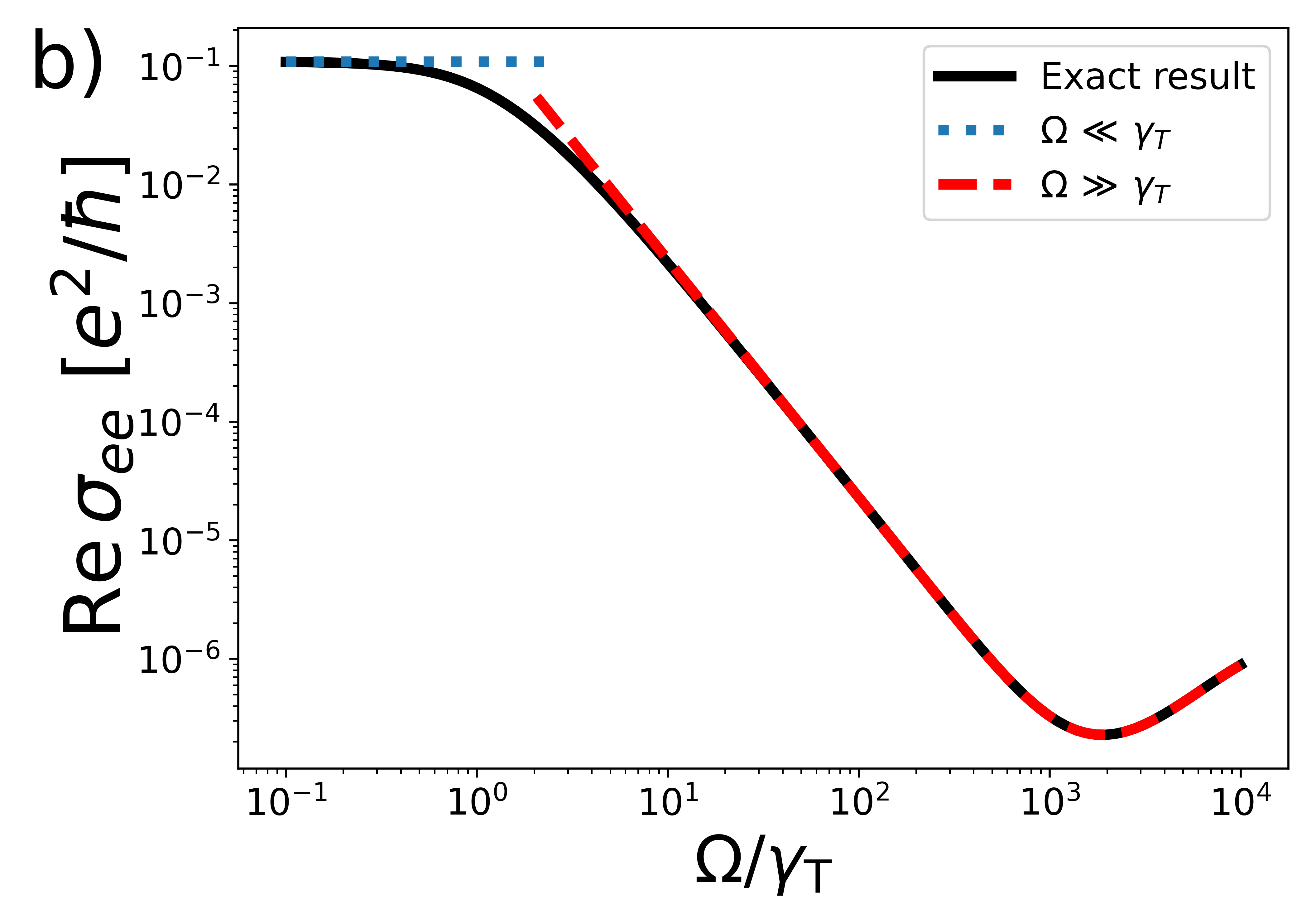}
\label{optical}
    \end{subfigure}
    \caption{ 
(a) \emph{ee} contribution to the \emph{dc} conductivity of a FL with impurities, $\sigma_{\rm ee}(0,T)$, as a function of $\gamma_T \taui$  
for
$\Omega_{\rm p}/\ef=0.6$ and $ \varepsilon_F\taui = 100$, where $\gamma_T$ is defined in Eq.~\eqref{gamma}. Left axis: $\see(0,T)$ in units of $e^2/\hbar$; right axis: $\see(0,T)$ normalized by the residual conductivity, $\sigma_\mathrm{i}(0)$.
Main panel: exact solution, Eq.~\eqref{scalingF2} for $\Omega=0$. 
Top and bottom insets: high- and low-temperature  asymptotic limits of $\see$, respectively. 
(b) Optical conductivity $\mathrm{Re}\,\sigma_{\rm ee}(\Omega,T)$ of an impurity-free FL, given by the $\taui\to \infty$ limit of Eq.~\eqref{scalingF2},  as a function of $\Omega/\gamma_T$, in units of $e^2/\hbar$ and for $T=0.1\Omega_{\rm p}=0.01\ef$. 
In both panels, the dotted (blue) and dashed (red) curves correspond to analytical results. \label{combined}}   
\end{figure}

 As announced, 
 the ratio of the $T$-dependent part of $\sigma$ to its residual value is at most of order $T^2/\ef^2$, and 
 the crossover between the $T^4|\ln T|$ and $T^2$ regimes occurs when $\gamma_T\taui\sim 1$. The \emph{ee} part of the dc conductivity
 is plotted in Fig.~\ref{combined}a as a function of 
$\gamma_T\taui$.

 Next is the optical conductivity in the impurity-free limit 
 ($\Omega\taui\to\infty$),
 when
  $\lambda=-i\Omega/\gamma_T$.
Since $\gamma_T\ll T$ in a FL, the hydrodynamic/collisionless crossover at $\Omega\sim\gamma_T$ lies entirely within the semiclassical regime $\Omega\ll T$. For
 $\gamma_T\ll \Omega\ll T $,
Eq.~\eqref{scalingF2}
reproduces the Drude-like tail of the optical conductivity: 
 $\R\see(\Omega,T)=4\pi e^2T^2\gamma_T/15 \ef \Omega^2\propto T^4 |\ln T|/\Omega^2$  \cite{Sharma:2021}.
For $\Omega\gtrsim T$, the semiclassical KE is not applicable.
 We found, however, that its  collisionless result can be extrapolated into this regime by making two replacements. First, the phase-space factor $T^2$ is replaced by $T^2 + \Omega^2/4 \pi^2$, as required by the first bosonic Matsubara-frequency rule \cite{maslov:2012}. This gives the function $\Phi(\Omega,T)$ in Eq.~\eqref{scalingF2}.
 Second, $\gamma_T$ is replaced by
$\tauee^{-1}(\Omega,T)=\pi(T^2+3\Omega^2/8\pi^2)L_{\rm C}(
\mathcal{E})/4\ef$. 
With these replacements, Eq.~\eqref{scalingF2} reproduces the result of Ref. \cite{Sharma:2021}, derived independently from the Kubo formula, throughout the collisionless regime.We regard this agreement as a consistency check rather than as an extension of the semiclassical KE to $\Omega\gtrsim T$.

 In addition, our result allows us to study the hydrodynamic regime, not accessible by perturbation theory in $\gamma_T/\Omega$.
 For $\Omega\ll \gamma_T$ ($|\lambda|\ll 1$), the small-$\lambda$-form of $\mathcal{F}$ implies that the conductivity approaches an $\Omega$-independent limit:  $\R\see(\Omega\to 0^+,T)=e^2/2 L_{\rm C}(T)$. 
The frequency dependence of  $\R\see(\Omega,T)$ is shown in Fig.~\ref{combined}b.

\paragraph*{Nearly critical FL.}
The KE for a nearly critical FL can also be derived within the Keldysh formalism 
(See Appendix~\ref{app:KE_SC})  .
Within the approximation that the nonequilibrium part of the retarded self-energy is neglected \cite{mahan:book}, the resulting equation has the same form as Eq.~\eqref{KE}, except that the driving term contains the renormalized Fermi velocity. In addition, all FL parameters entering the collision integral are renormalized. The scattering kernel is obtained from the static Hertz-Millis interaction in Eq.~\eqref{Ueff} 
as
\bea
W(q)=\frac{2\pi Z^4 g^2}{(q^2+\xi^{-2})^2}.\label{WHM}
\eea
The factor of \(Z^4\), introduced in Ref.~\cite{lifshitz:1980} on phenomenological grounds, follows naturally from the quasiparticle ansatz for the lesser Green's function,
$G^{-+}(\omega,\bk)=2\pi i Z\,\delta(\omega-Z\ve_\bk)\,f_\bk$,
since the Keldysh collision integral for \emph{ee} scattering contains convolutions of four Green's functions. The conductivity is still given by Eqs.~\eqref{scalingF} and \eqref{scalingF2}, provided one replaces
\[
\frac{1}{\tauee(\mathcal{E})}\to \frac{1}{\tauee^*(\mathcal{E})}
=\frac{\mathcal{E}^2L_{\rm HM}(\mathcal{E})}{2\ve_{\rm HM}},
\qquad
\ve_{\rm HM}=\frac{2\pi^3}{g^2\ef^*}\left(\frac{\vf^*}{Z\xi}\right)^4,
\]
and \(\ef\to \ef^*\) everywhere. The high-frequency limit then yields
\bea
\R\see(\Omega,0)=\frac{1}{160\pi^4} e^2 g^2 \left(\frac{Z\xi}{\vf^*}\right)^4\Omega^2 \ln\frac{\mathcal{E}_{\rm FL}}{|\Omega|}.
\eea
Since \(\vf^*=Z\vf\), the final result depends only on bare quantities, as required by the Ward identity, which is implemented explicitly in the Kubo approach of Ref.~\cite{Gindikin:2025}. In the hydrodynamic regime,
\bea
\R\see(\Omega\to 0^+,T)=\frac{\pi^2}{2} e^2 \frac{1}{g^2\ef^{*2}}\left(\frac{\vf^*}{Z\xi}\right)^4\frac{1}{\ln(\mathcal{E}_{\rm FL}/T)}.
\eea
The crossover between these two limits, as well as the intermediate regime $\gamma_T\ll\Omega\ll T$, is captured qualitatively by Eq.~\eqref{inter}. The extrapolation of these  results to the non-FL regime has already been described above. Although this procedure, taken literally, 
was shown to yield the wrong sign of $\R\see$ for $\Omega\gg T$ \cite{Gindikin:2025}, it nevertheless reproduces the correct scaling form.

\paragraph*{Acknowledgments.} We are grateful to A. V. Chubukov, Y. Gindikin, and S. Hartnoll for stimulating discussions. The work of T.K. and D.L.M. was supported by the National Science Foundation  via DMR-2224000. The work of 
V.I.Y. was supported by the Basic Research
     Program at the HSE University (HSE-BR-2025-57).

The data that support the findings of this article are openly
available
\cite{kiliptari2026data}.
\appendix

\section{Comparison of semiclassical and quantum-interference contributions}
\label{app:QI}

The \emph{ee} contribution to the conductivity considered in this paper is of purely semiclassical origin. It arises because \emph{ee} interaction modifies the electron energy distribution and requires a non-parabolic dispersion. On the other hand, quantum-interference mechanisms also give rise to temperature-dependent corrections to the conductivity, and are present even for a parabolic dispersion. In this section, we briefly compare the magnitudes of the semiclassical and quantum-interference corrections, and identify the temperature window where the semiclassical correction can dominate over the quantum-interference one. Such a comparison is necessary because the semiclassical correction discussed in the paper never exceeds the conductance quantum, $\sigma_0=2e^2/h$, whereas quantum-interference corrections generically exceed $\sigma_0$ in 2D.

In 2D, weak localization leads to a logarithmic temperature dependence of the conductivity, $|\delta\sigma_\mathrm{WL}|\sim \sigma_0\ln(L_{\phi}/\vf\taui)$, where $L_\phi$ is the phase-breaking length. This correction is suppressed by a relatively weak magnetic field, which does not lead to substantial classical magnetoresistance. In what follows, we assume that the weak-localization correction is already suppressed.

Next is the quantum-interaction correction, which arises from quantum interference between \emph{ei} and \emph{ee} scatterings \cite{altshuler:1985,Zala:2001}. Its temperature dependence is different in the diffusive ($T\taui\ll 1$) and ballistic ($T\taui\gg 1$) regimes.

In the diffusive regime, $\delta\sigma_\mathrm{int}\sim \sigma_0 g_\mathrm{D}\ln (1/T\taui)$, where $g_\mathrm{D}$ is the coupling constant, with a magnitude and sign depending on the interaction amplitudes in the singlet and triplet channels \cite{altshuler:1985}. For $|g_\mathrm{D}|\sim 1$, $\delta \sigma_\mathrm{AA}$ is larger than $\sigma_0$, while the semiclassical correction is smaller than $\sigma_0$ at low $T$; the interaction correction dominates.

In the ballistic regime ($T\taui\gg 1$), the quantum-interaction correction becomes linear in $T$: $\delta\sigma_\mathrm{int}\sim \sigma_0 g_\mathrm{B} T\taui$, where $g_\mathrm{B}$ is a coupling constant different from $g_\mathrm{D}$ \cite{Zala:2001}. For $|g_\mathrm{B}|\sim 1$, $|\delta\sigma_\mathrm{int}|\sim \sigma_0 T\taui\gg \sigma_0$. Taken at face value, this implies that the quantum-interaction correction wins in the ballistic regime as well. However, the result for $\delta\sigma_\mathrm{int}$ quoted above applies to the case of point-like impurities. Since the ballistic limit of the quantum-interaction effect arises due to backscattering of electrons by a Friedel oscillation around a single impurity, its magnitude is proportional to the $2\kf$ component of the disorder correlator. For smooth disorder such a component is suppressed, and may be even exponentially small in semiconductor heterostructures with remote donors \cite{gornyi:2004}.

Although we also consider point-like impurities in this paper, an extension to long-range disorder is straightforward: $\taui$ needs to be replaced by the transport time, $\taui^\mathrm{tr}$. Given that the ballistic quantum-interaction correction is suppressed for long-range disorder and the weak-localization correction is suppressed by the magnetic field, we therefore expect the correction considered here to be observable, and potentially dominant over the ballistic quantum-interaction correction. For a FL with Coulomb interaction, the corresponding temperature window is given by
\bea
1/\taui^\mathrm{tr}\ll T\ll \Lambda_\mathrm{C},\label{cross_app}
\eea
where $\Lambda_\mathrm{C}$ is the upper cutoff of the Coulomb model.

The quantum-interaction correction is enhanced near a QCP. For example, near a ferromagnetic QCP, $\delta\sigma_\mathrm{int}$ behaves as $\ln^2 T$ and $T^{1/3}$ in the diffusive and ballistic regimes, respectively \cite{Paul:2005}. However, the crossover between the two regimes is now at
\bea
T^*\sim \frac{1}{g\ve_F\taui^2}\frac{1}{\taui},
\eea
where $g$ is the coupling constant (of units of energy) of the HM model, defined in Eq.~\eqref{Ueff}. For a sufficiently clean metal, it is realistic to assume that $g\ve_F\taui^2\gg 1$, and hence $T^*\ll 1/\taui$. As in the FL case, invoking smooth disorder strongly suppresses the ballistic quantum-interaction correction. Therefore, near a QCP the lower boundary of the temperature window in Eq.~\eqref{cross_app} can be reduced parametrically below $1/\taui$, down to the scale of $T^*$, and the temperature window where the semiclassical correction can be expected to dominate becomes wider:
\bea
\frac{1}{g\ve_F\left(\taui^\mathrm{tr}\right)^2}\frac{1}{\taui^\mathrm{tr}}\ll T\ll \Lambda_\mathrm{HM},
\eea
where $\Lambda_\mathrm{HM}$ is the upper cutoff of the HM model.

\section{Origin of the suppressed current relaxation in an isotropic Fermi liquid}
\label{app:suppression}

For an isotropic FL, the suppression arises because the current density for an electron with energy $\ve_\bk$ relative to $\ef$ can be decomposed into the conserved part, proportional to the momentum, and the non-conserved part, proportional to the thermal current:
\bea
\mathbf{j}=-e\bv_\bk=-e\vf\frac{\bk}{k}
-e\frac{\vf'}{\vf}\ve_\bk\frac{\bk}{k},
\eea
where $\vf'=\partial_k v(k)\vert_{k=\kf}$. Because the correlation function of the non-conserved part contains a factor of $\ve_\bk^2$, the current relaxation is suppressed.

We stress that this mechanism is different from the one operative in a 2D FL, where odd angular harmonics of the distribution function relax as $T^4$ \cite{levitov:2019}. The latter effect describes free decay rather than driven transport. In the driven case, the field-induced nonequilibrium distribution has nontrivial energy dependence. A clear signature of the difference between the two mechanisms is that the Sommerfeld-like suppression occurs in both 2D and 3D \cite{Sharma:2021,Goyal:2023}.

\section{Reduction of the integral kinetic equation to a differential form}
\label{app:reduction}

Expanding
\[\Delta \bv \equiv \bv_{\mathbf{k}}+\bv_{\mathbf{p}}
-\bv_{\mathbf{k}-\mathbf{q}}-\bv_{\mathbf{p}+\mathbf{q}}\]
in Eq.~\eqref{source} to second order in $\omega/\ef$, $\varepsilon_{\mathbf{k}}/\ef$, and $\varepsilon_{\mathbf{p}}/\ef$ yields:
\begin{widetext}
\begin{align}
\frac{\Delta v_x}{v_F}
=\left[-\frac{\omega}{\varepsilon_F}
+ \frac{(\varepsilon_{\mathbf{k}}-\omega)\omega}{\varepsilon_F^2}\right]
\cos\phi_{\mathbf{k}}
-\left[-\frac{\omega}{\varepsilon_F}
+ \frac{(\varepsilon_{\mathbf{p}}+\omega)\omega}{\varepsilon_F^2}\right]
\cos\phi_{\mathbf{p}}
+ \cos\phi_{\mathbf{q}}\,
\frac{\varepsilon_{\mathbf{p}}-\varepsilon_{\mathbf{k}}+2\omega}{\varepsilon_F}
\frac{q}{k_F},\nn
\end{align}
\end{widetext}
where $\phi_\mathbf{n}$ is the azimuthal angle of the vector $\mathbf{n}$.

The angular integrations over the angle between $\bk$ and $\bq$ ($\phi_{\mathbf{k}\mathbf{q}}$), and between $\bp$ and $\bq$ ($\phi_{\mathbf{p}\mathbf{q}}$) are performed using the energy-conserving delta functions, e.g.,
\[\int \frac{d\phi_{\mathbf{k}\mathbf{q}}}{2\pi}\,
\delta(\omega-v_F q\cos\phi_{\mathbf{k}\mathbf{q}})
=\frac{1}{\pi v_F q}.
\]
As a result, terms proportional to $\cos\phi_{\mathbf{k}}$, $\cos\phi_{\mathbf{q}}$, and $\cos\phi_{\mathbf{p}}$ generate contributions scaling as $\cos\phi_{\mathbf{k}}/(\pi v_F q)^2$ multiplied by powers $(\omega/v_F q)^0$, $(\omega/v_F q)^1$, and $(\omega/v_F q)^2$. Only the first type of terms produces a logarithmically large contribution from the $q$ integration; all others are neglected.

Carrying out the remaining $\omega$ and $\varepsilon_{\mathbf{p}}$ integrations and expanding the density of states, we obtain
\begin{equation}
I(\bk)=\frac{-\pi\cos\phi_\bk\,\taui\, L_{\rm C}(T)}{6\pi\ef^2}\left[
\ve_\bk(\ve_\bk^2+\pi^2T^2)-\frac{3}{4\ef}(\ve_\bk^2+\pi^2T^2)^2\right].\nn
\end{equation}
As noted above, we keep only the odd-in-$\ve_\bk$ contribution, given by the first term.

We now analyze Eq.~\eqref{Sterm}, which contains four terms. The second term, proportional to $\bv_{\bk-\bq}-\bv_{\bk}$, vanishes to second order in the small parameters specified above. The third term, containing $\bv_{\bp}$, does not generate a large logarithm and is neglected. We therefore retain only the first and last terms. Integrating over the angles and momentum transfer, we obtain
\bea
S(\bk)&=&\cos\phi_\bk\int d\ve_\bp \int d\omega\,\nu\!\left(1+\frac{\ve_\bp}{\ef}\right)
K(\omega,\ve_\bk,\ve_\bp)\nn\\
&&\times\left[F(\ve_\bk)-F(\ve_\bk-\omega)
-\frac{\omega}{\ef} F(\ve_\bp+\omega)\right]\nn\\
&\equiv&\cos\phi_\bk(N-S_1-S_2),\nn
\eea
where
\[K(\omega,\ve_\bk,\ve_\bp)=
\frac{1-n(\ve_\bk-\omega)}{1-n(\ve_\bk)}\,n(\ve_\bp)(1-n(\ve_\bp+\omega)).\]
The first term in the bracket is the ``scattering-out'' part of the collision integral ($N$), while the remaining two form the ``scattering-in'' part ($S_1+S_2$). Carrying out the integrations over $\ve_\bp$ and $\omega$ in the $N$ term, we obtain $N = N_{\rm e} + N_{\rm o}$, where
\bea
N_{\rm e}&=&\frac{L_{\rm C}}{4\pi\ef}(\ve_\bk^2+\pi^2 T^2),\nn\\
N_{\rm o}&=&-\frac{L_{\rm C}}{12\pi}\frac{\ve_\bk}{\ef^2}(\ve_\bk^2+\pi^2 T^2).\nn
\eea
Integrating over $\ve_\bp$ in the $S_1$ term gives
\bea
S_1=\cosh(\ve_\bk/2T)\frac{L_{\rm C}}{2\pi\ef}\int du\, M(\ve_\bk-u)\,H(u),\nn
\eea
where $M = M_{\rm e} + M_{\rm o}$, $H = H_{\rm e} + H_{\rm o}$,
\bea
M_{\rm e}(\epsilon)&=&\frac{\epsilon}{2\sinh(\epsilon/2T)},\quad
M_{\rm o}(\epsilon)=-\frac{\epsilon^2}{4\ef\sinh(\epsilon/2T)},\nn
\eea
and
\begin{equation*} 
H_{\rm e,o}(\epsilon)=F_{\rm e,o}(\epsilon)/\cosh(\epsilon/2T).\end{equation*}

The term $S_2$ does not contribute to the odd part of $F$ and is thus discarded.

To obtain an equation for the odd part of $H$, we isolate the even parts of $N$ and $M$ while keeping the odd parts of $H$. The $H_{\rm e}M_{\rm o}$ term in $S_1$ is neglected since $F_{\rm o}\gg F_{\rm e}$. Introducing $Y_0=J_0/\cosh(\ve/2T)$, we arrive at
\bea
\left[N_\mathrm{e}(\ve)+\frac{1}{\taui}\right]H_\mathrm{o}(\ve)
-\frac{L_{\rm C}}{2\pi\ef}\int du\,M_\mathrm{e}(\ve-u)\,H_\mathrm{o}(u)=Y_\mathrm{o}(\ve).\nn
\eea
To arrive at Eq.~\eqref{KEmain}, we introduce the dimensionless variable $\xi=\ve/T$ and the rescaled functions
\bea
M_{\rm e}&=&T\mathcal{M}_{\rm e},\quad
N_{\rm e}=\frac{T^2L_{\rm C}}{4\pi\ef}\mathcal{N}_{\rm e},\nn\\
Y_{\rm o}&=&\frac{\lambda\taui T^3 L_{\rm C}^2}{16\ef^2}\mathcal{Y}_{\rm o},\quad
H_{\rm o}=\mathcal{H}_{\rm o}/T.\nn
\eea

\paragraph{Asymptotic form of $\mathcal{F}$ for $\R\beta\gg 1$.}

In this limit, the Green's function in Eq.~\eqref{Gf2} is strongly peaked near $s=s'$, where it can be approximated as
\bea
\mathcal{R}_\beta(s,s')\approx\frac{1}{4\beta^3}\exp\!\left[-\frac{\beta|s-s'|}{1-s^2}\right].\nn
\eea
Substituting this into Eq.~\eqref{functionF} and evaluating the pre-exponential factor at $s=s'$, we obtain
\bea
\mathcal{F}(\R\beta\gg 1)=-\frac{4\pi}{15\beta^2},\nn
\eea
equivalent to the form $\mathcal{F}(|x|\gg 1)=-4\pi/15x$ quoted after Eq.~\eqref{functionF}.

\paragraph{Asymptotic form of $\mathcal{F}$ for $|\beta|\to 1$.}

According to Eq.~\eqref{functionF}, the prefactor of the conductivity has an imaginary part. To obtain the real part of $\sigma$, one therefore needs both the real and imaginary parts of $\mathcal{F}$. In this limit we expand the fundamental solutions $P^{\pm\beta}_1(s)$ to second order in $\beta-1$, taking into account that $\mathcal{F}$ contains the factor $\beta^2-1$ through $\mathcal{C}_\beta$ in Eq.~\eqref{Gf2}. This yields
\bea
\mathcal{F}(|\beta|\to 1)=-\frac{\pi}{6}+\frac{\pi\lambda}{8},\nn
\eea
equivalent to the form $\mathcal{F}(|x|\ll 1)=-\pi/6+\pi x/8$ quoted after Eq.~\eqref{functionF}.

\section{Kinetic equation for a strongly coupled Fermi liquid}
\label{app:KE_SC}

To first order in gradients, the Dyson equation for the lesser component of the matrix Green's function in the presence of a weak, uniform, oscillatory electric field $\mathbf{E}(t)=-\partial_t\mathbf{A}(t)$ reads \cite{rammer:1986}
\begin{widetext}
\bea
&&\bigl[1-\partial_\omega\R\Sigma_0^R(p)\bigr]\partial_t\delta G^<(p,t)
+\bigl[-e\,\bv_{\bk}\cdot\bE(t)
+\partial_t\R\delta\Sigma^R(p,t)\bigr]\partial_\omega G_0^<(p)\label{linear-Gless-homogeneous-app}\\
&&-\partial_\omega\Sigma_0^<(p)\,\partial_t\R\delta G^R(p,t)
+\partial_t\delta\Sigma^<(p,t)\,\partial_\omega\R G_0^R(p)=
i\Sigma^<(p,t)G^>(p,t)-i\Sigma^>(p,t)G^<(p,t),\nn
\eea
\end{widetext}
where $p\equiv(\bk,\omega)$, and $X_0$ and $\delta X=X-X_0$ denote the equilibrium and nonequilibrium parts of $X$, respectively. Equation~\eqref{linear-Gless-homogeneous-app} is local in $p$, as appropriate in the semiclassical limit ($\Omega\ll T$).

The distribution function $f_\bk$ is introduced through the quasiparticle ansatz
\bea
G^<(p)&=&2\pi i Z\,\delta(\omega-\ve_\bk^*)\,f_\bk,\nn\\
G^>(p)&=&-2\pi i Z\,\delta(\omega-\ve_\bk^*)\,(1-f_\bk).\label{qp-app}
\eea
In the Eliashberg case considered here, the self-energy does not depend on $\bk$, and thus $\ve^*_\bk=Z\ve_\bk$. The nonequilibrium part of $G^<$ comes from two sources: the variation of $f_\bk$ and the shift of the quasiparticle pole,
\bea
\delta G^<=2\pi i\left[\bigl\{Z\delta f_\bk+\delta Z\,n_\bk\bigr\}\delta(\omega-\ve_\bk^*)
-Z\delta'(\omega-\ve_\bk^*)n_\bk\delta\ve_\bk^*\right],\nn
\eea
where
\bea
\delta\ve_\bk^*=e\bv_\bk^*\cdot\mathbf{A}(t)+Z\partial_\omega\R\delta\Sigma^R(\ve_\bk^*,t),\label{deltae-app}
\eea
with $\bv^*_\bk=Z\bv_\bk$. The $\delta'(\omega-\ve_\bk^*)$ terms in Eq.~\eqref{linear-Gless-homogeneous-app} cancel provided Eq.~\eqref{deltae-app} holds.

The first two terms in the second line of Eq.~\eqref{linear-Gless-homogeneous-app} contain the principal-value parts of $G^R$ and are thus less singular than the $\delta$-function terms in the first line. Discarding the principal-value parts and integrating the resulting equation over $\omega$, we obtain a KE for $f_\bk$:
\bea
\partial_t\delta f_{\bk}
&+&\Bigl(-e\,\bv^*_{\bk}\cdot\mathbf{E}(t)
+Z\partial_t\R\delta\Sigma^R(\ve_{\bk}^*,t)\Bigr)n_\bk'=-I,\nn\\
&&I=\frac{1}{2\pi}\int d\omega\left(\Sigma^> G^<-\Sigma^< G^>\right).
\label{eq:KE1-app}
\eea
If the term $Z\partial_t\R\delta\Sigma^R(\ve_{\bk}^*,t)$ in Eq.~\eqref{eq:KE1-app} is neglected \cite{mahan:book}, the left-hand side reduces to the canonical form with a renormalized velocity in the driving term. If the nonequilibrium part of the retarded sector is also neglected in the collision integral, its \emph{ee} part reduces to the canonical form in the standard way \cite{rammer:1986}.

We now illustrate the origin of the factor $Z^4$ in Eq.~\eqref{WHM} using the scattering-out term as an example:
\bea
I^{\rm out}_{\rm ee}=\int\frac{d\omega}{2\pi}\left(\Sigma_0^>\delta G^<-\delta\Sigma^> G^>_0\right),\label{out-app}
\eea
where the one-loop self-energy is given by
\bea
\Sigma^{\alpha}(\omega)=i\int_{\bq,\omega'}U^{\alpha}_{\rm HM}(q,\omega')G^{\alpha}(\bp+\bq,\omega+\omega'),\,\,\alpha=\{>,<\}.\nn
\eea
In the FL regime, $U^\alpha_{\rm HM}(q,\omega)\approx U^2_{\rm HM}(q,0)\Pi^\alpha(q,\omega)$, where
\bea
\Pi^{>,<}(q,\omega')=-i\int_{\bp,\epsilon}G^{>,<}(\bp+\bq,\epsilon+\omega')G^{<,>}(\bp,\epsilon)\nn
\eea
is the polarization bubble.

Since each term in Eq.~\eqref{out-app} contains a product of four Green's functions from Eq.~\eqref{qp-app}, $U^2_{\rm HM}(q,0)$ acquires a factor of $Z^4$, in agreement with Ref.~\cite{lifshitz:1980}.

\bibliography{dm_references}
\end{document}